# Uncertainty Quantification in Classical Molecular Dynamics

## Shunzhou Wan[1], Robert C. Sinclair[1], and Peter V. Coveney[1,2]*

*[1]Centre for Computational Science, University College London, Gordon Street, London WC1H 0AJ, UK,*
*[2]Institute for Informatics, Science Park 904, University of Amsterdam, 1098 XH Amsterdam, Netherlands,*

*https://orcid.org/0000-0002-8787-7256 (PVC)*
*https://orcid.org/0000-0001-7192-1999(SW)*
*https://orcid.org/0000-0003-0418-6415(RCS)*

## Abstract

Molecular dynamics simulation is now a widespread approach for understanding complex systems on the atomistic scale. It finds applications from physics and chemistry to engineering, life and medical science. In the last decade, the approach has begun to advance from being a computer-based means of rationalising experimental observations to producing apparently credible predictions for a number of real-world applications within industrial sectors such as advanced materials and drug discovery. However, key aspects concerning the reproducibility of the method have not kept pace with the speed of its uptake in the scientific community. Here, we present a discussion of uncertainty quantification for molecular dynamics simulation designed to endow the method with better error estimates that will enable the method to be used to report actionable results. The approach adopted is a standard one in the field of uncertainty quantification, namely using ensemble methods, in which a sufficiently large number of replicas are run concurrently, from which reliable statistics can be extracted. Indeed, because molecular dynamics is intrinsically chaotic, the need to use ensemble methods is fundamental and holds regardless of the duration of the simulations performed. We discuss the approach and illustrate it in a range of applications from materials science to ligand-protein binding free energy estimation.

## Introduction

*Author for correspondence: p.v.coveney@ucl.ac.uk.

†Centre for Computational Science, University College London

Gordon Street, London WC1H 0AJ, UK

Computational methods offer a route to understand and predict structure, dynamics and thermodynamics of molecular systems. A large fraction of these are primarily based on molecular dynamics (MD) simulations, first developed in the late 1950s[1]. These methods, however, find their applications only in a limited segment of industry. Only in the past decade have they started to be used in materials manufacturing[2] and more recently in the pharmaceutical industry[3]. In comparison to macroscopic modelling methods such as the finite element method (FEM) or computational fluid dynamics (CFD), which enjoy widespread use in engineering applications, MD methods remains primarily confined to academic research, largely due to their lack of reproducibility, and limited accuracy, as well as the frequent long duration and computational expense required for their use. Accuracy, precision and reproducibility are essential in any method which is to be relied upon for taking actionable decisions and thus to become valuable in diverse applications, including *inter alia* industrial and clinical contexts. For that, we need uncertainty quantification (UQ), verification and validation (V&V), or VVUQ. But while careful control of uncertainty is the mainstay of weather forecasting, along with many branches of engineering and applied mathematics, it is rather rarely performed in disciplines such as physics and chemistry where much time is spent investigating matter at shorter length and time scales than the macroscopic ones of direct concern in many real world situations. The purpose of the present paper is to assess the use of uncertainty quantification in the field of molecular dynamics simulations and its relationship to the reproducibility of the method. We do not provide a comprehensive review of molecular dynamics applications. We shall spend much less time looking at verification and validation, which are respectively concerned with whether our computer programs are solving the correct equations and how well their output agrees with experimental observations[4].

Computational predictions are increasingly being used to predict outcomes and provide recommendations in a variety of industrial and policy-making contexts. One of the central aims of UQ is to facilitate decision making[5]; this is enabled by providing probabilistic statements about quantities of interest, in a timely fashion, and helping decision makers to decide what actions to take that maximise the likelihood of a desired outcome. Ensemble methods for probabilistic weather prediction have become a routine part of the practice of weather forecasting[6]. Simulation results have been used to help governments and non-governmental organisations to make decisions as to how to help plan ahead for mass movements of refugees[7]. Accurate and rapid binding free energy predictions[8] are starting to impact decision making in the pharmaceutical industry and clinical settings[9,10]. A very recent and topical example are predictions emanating from simulations of an epidemiological model by Ferguson et al. which were used to guide UK government policy in addressing the COVID-19 pandemic[11–13].

The steady increase of computational power permits investigations of a large diversity of models over increasing length and time scales. This makes it even more essential to systematically assess the reliability and reproducibility of the methods used and the results generated, as such large-scale applications involve a vast amount of computational time and human effort. Uncertainty quantification, along with validation and verification, ensure that the results are reliable and reproducible while conferring greater confidence on predicted outcomes. Without UQ, the usefulness and value of simulations is diminished, and the confidence in computational results degraded. Science and engineering manifestly advance faster when there is less time wasted on pursuing false leads.

Given the increasing application of scientific computation in critical decision making, uncertainty quantification has been the subject of broader attention in a range of domains[14]. In computational physics and computational chemistry, for example, there are well documented cases in the literature assessing the reliability with which various quantities of interest can be computed. In ensembles of exchange-correlation functionals for density-functional theory calculations, large error bars have been observed on energy differences at different states[15], in reaction energies of specific chemical systems[16], within reaction kinetics of surface transition states[17], and estimated free energies of activation[18]. While the uncertainties in a model's parameters contribute to the uncertainties in the predictions, caution is needed when estimating prediction uncertainty based on parameter uncertainty inflation[19]. In some instances, prediction errors due to model inadequacy can be handled by statistical correction of predictions, which may provide a reliable uncertainty measure[20]. Various methods have been developed for the estimation of prediction uncertainty, such as bootstrap-based methods, Gaussian process regression, neural networks and deep learning ensembles[21–23]. Gaussian process regression has been employed to identify particular calculations within a given data set for which the uncertainties exceed a given threshold[24,25]. The data points flagged up can be investigated with refined models and more accurate methods that may be added back to the data set if their uncertainties then fall below the threshold concerned. Such validation and updating can improve the models and enhance the quality of subsequent predictions[24].

One major application of MD simulation that we consider is the prediction of the binding affinity of a lead compound or drug candidate with a protein target, of major relevance in drug discovery and personalised medicine. That target may be respectively either a generic protein or a sequence specific variant, reflecting the fact that individuals respond differently to a given drug based on their genetic make-up. The binding affinity, also known as the free energy of binding, is the single most important initial indicator of drug potency, and the most challenging to predict[26,27]. Another case we look at here is how one seeks to make actionable predictions in support of advanced materials discovery. The guiding principle in all instances is to seek to make reproducible predictions.

Reproducibility is an intrinsic feature of the scientific method, whether experimental or computational. Scientific methods should yield the same results in a statistical sense regardless of who performs them. Indeed, the lack of reproducible results in the published literature is of current concern in the wider scientific community[28–30]. It should be noted that chaotic dynamical systems exhibit extreme sensitivity to initial conditions, making accurate predictions impossible and one-off observations largely unreproducible even though their underlying dynamics is deterministic. Molecular dynamics is a case in point for which these issues need to be addressed. To restore the predictive power of the scientific method to such systems scientists contend that, while the accuracy with which we can simulate an individual chaotic process is severely limited, accuracy in an averaged statistical and reproducible sense may still be possible[31]. The purpose of the present paper is to assess the reproducibility and intrinsic uncertainty of molecular dynamics simulation. We illustrate the issues by way of some examples drawn from materials and life sciences. The discussion is, of course, applicable to all areas of classical molecular dynamics.

## Sources of Error in Classical Molecular Dynamics

There are two sources of error accruing in MD simulations, due to systematic and random sources. In order to get a full grip on uncertainty in MD simulations, one needs to be able to identify both. Systematic errors originate in things like the imperfect design, parameterisation, conduct and/or analysis of a study, which result in an estimate of a property deviating consistently from its true value. Random variation—also called system noise, aleatoric or stochastic error—on the other hand, is caused by the intrinsically chaotic nature of classical molecular dynamics and produces apparently random deviations from the notionally "true" value of an observable. It should be noted that in some cases no consensus definitions for the "true value" may exist; more precisely, differences in definitions from modelling and experimental studies can then contribute to observed errors in validation studies. The glass transition temperature and the setting of cements, for example, have a number of different operational definitions although only some of them are endorsed as accepted scientific or engineering standards. The simulations should proceed from a statistical-mechanical ensemble corresponding to the experimental conditions and properties calculated from expectation values may then be compared with their corresponding experimental counterparts. Quantifying systematic errors requires first bringing the random components contributing to the errors under control.

### Systematic Errors

Systematic errors are introduced by inaccuracies inherent to the system investigated and within the measurement method performed. They come from the assumptions and approximations made when a theory is applied, a model is constructed, or a process is mimicked by the simulation of a real-life problem. In constructing a model, there are many choices to be made, including: which degrees of freedom are to be modelled explicitly, what components are to be excluded, the kind of interactions between the components, what boundary conditions are to be used, and so on. As Michael Levitt has stated: "the art is to find an approximation simple enough to be computable, but not so simple that you lose the useful detail"[32]. In principle, a higher level of resolution should produce more accurate predictions than a lower level one, although in practice it is not always the case because of the quality of the theory employed, a fortuitous cancellation of errors, or the way that the methods are implemented which may not be fully verified. It is not uncommon that simple methods outperform more complicated methods in the simulation field, as have been seen in the blind SAMPL free energy prediction competitions[33–35]. In drug discovery approaches, a ligand-protein model with explicit water molecules is usually better than one with implicit water. These choices all affect the outcome of a simulation, usually in a deterministic way. Biases in the interaction parameters chosen to represent the system can significantly influence the results; for example, different protein force fields favour different secondary structure types[36,37], populating either helical or sheet-like structures within independent simulations. When the cause of such systematic errors can be identified, it can be reduced or even eliminated, as shown for example in recent simulations with state-of-the-art force fields[38].

The implementation of the model in an appropriate MD engine and the calibration of the engine can also influence the results. The thermodynamic conditions, such as constant volume or pressure in a closed system, must be specified. Multiple factors need to be carefully considered in the preparation of the molecular models, such as choice of force field, protonation and tautomeric states, buffer conditions, use of restraints and constraints, thermostat and barostat, free energy estimator, and finite machine representation of floating point numbers[39,40]. A few operational parameters need to be fine-tuned, including those for temperature and pressure couplings, for the calculation of long-range interactions, for the time step(s) used within the integration algorithms, and so on. Other factors, such as the introduction of numerical integrators and the accumulation of rounding error[31], may also lead to systematic patterns of error. Moreover, it is entirely possible that molecular dynamics may manifest a pathology which we recently discovered in the simulation of simple chaotic systems on digital computers[41]. It is caused by the limitations of the IEEE floating point numbers in describing the statistical behaviour of systems with such exquisite sensitivity: ensemble averages, designed to address random errors, also contribute substantial systematic errors to predicted properties[41]. The pathology cannot be mitigated by increasing the precision of these numbers.

*Random Errors*

Given the extreme sensitivity of Newtonian dynamics to initial conditions, two independent MD simulations will sample the microscopic states with different probabilities no matter how close the initial conditions used[42]. The difference thus produced in two simulations introduces a variation in results that can often be larger than the quantity of interest, making the results practically useless. Large hysteresis can be observed in cases where adequate sampling of all relevant conformational substates is not achieved[43,44]. It should be noted that a seemingly low standard deviation does not guarantee convergence; it can appear when simulations remain trapped in a single energy well[45].

The impact of the chaotic nature of molecular dynamics has not been widely recognised in the molecular dynamics field. Most accounts give surprisingly short shrift to it, a notable exception being the recent book by Leimkuhler and Matthews (2015)[31], albeit it does not address the connected issue of uncertainty quantification and the estimation of thermodynamic quantities. Extensive studies we have performed in recent years show that molecular dynamics systems indeed exhibit extreme sensitivity to initial conditions[46–49]. From our investigations, we observe that the properties one computes from molecular dynamics trajectories appear *superficially* to be described by a Gaussian random process (GRP) with a normal distribution denoted by $N(\mu, \sigma^2)$, characterised by a $\mu$ and standard deviation $\sigma$ (the square root of the variance in the data). Note, however, that a normal distribution cannot be assumed and in fact there are frequently significant deviations from such statistics in nonlinear dynamical systems of which molecular dynamics is an excellent example[50].

In a recent study, for example, the effect of box size on simulations of protein dynamics in water was reported[51]; the authors reported that calculated values of various properties, such as the stabilities of the unliganded and liganded states of human hemoglobin and the density or number of hydrogen bonds per water molecule, changed systematically with an increase in box size; the authors maintained that a surprisingly large box of 150 Å was required to obtain meaningful results. Although at first sight this dependency on the box size appears to be an example of a systematic error in the simulation, it is in fact caused by a lack of reproducibility in the study which becomes manifest when random errors are taken into account[52]. Indeed, the ensuing debate[51–54] highlights the importance of setting up systems correctly for simulation and, more importantly, applying ensemble approaches to get statistically significant results. As we noted above, without first handling the stochastic errors, it is not possible to assess correctly the nature and magnitude of the systematic errors, and to interpret findings convincingly.

## Uncertainty Quantification in Molecular Dynamics Simulations

Although it was recognised more than two decades ago that one-off classical molecular dynamics simulations do not generate consistent protein conformations[55,56], systematic investigation as to how to make these calculations reproducible had not been performed until recently. Considerable effort has been invested in the development of so-called "enhanced sampling protocols" in order to accelerate phase space sampling, their purpose being to make computed properties more reliable by demonstrating more rapid "convergence" of computed properties. These enhanced sampling protocols accelerate molecular dynamics to overcome high energy barriers using methods such as bias potential approaches[57,58], Markov models[59], orthogonal space random walk[60], self-guided Langevin dynamics[61], and Hamiltonian replica exchange[62]. However, in all these cases it is quite impossible to calculate the (equilibrium or other) probability distribution function from one-off simulations, against which expectation values would be calculated; instead, expectation values of various observable are reported. Ensembles of such enhanced sampling simulations shows that there is significant variance between the expectation values computed from individual replicas[63].

Indeed, ensemble-averaging is not just a practical consideration invoked in the repertoire of uncertainty quantification methods. When molecular dynamics is used, as it frequently is, to estimate thermodynamic properties, such as the free energy of a system, it should be recalled that the connection between microstates (generated by individual MD simulation trajectories) and thermodynamic properties is achieved using ensemble averages. This is true whether the system is in or out of equilibrium. The very common resort to perform so-called "long time averages" of a single microsystem appeals to the ergodic theorem, which is in fact only valid for long times at which this time average should converge to the ensemble average. In reality, that time interval would need to be of the order of a Poincaré recurrence time—a truly astronomical epoch—for the equality to hold. In practice, it is taken to be as long as authors deem to be reasonable; and, compounding this, we must face the fact, mentioned above, that the accuracy of these long duration trajectories is severely limited.

In the ergodic hierarchy of dynamical systems, those which approach and reach equilibrium must be at least mixing[42]. Mixing systems are ergodic, but the converse is not true. Mixing systems exhibit the tell-tale property of dynamical chaos: neighbouring trajectories, no matter how close, diverge exponentially, at a rate given by a Lyapounov exponent. The point is that we are dealing with two levels of description and those wedded to trajectories and Newtonian mechanics think only in terms of one, the trajectories which follow Newton's equations of motion. To understand the concept of equilibrium, one must work with probability distributions (computed from ensembles) which obey Liouville's equation. Ergodic theory is about the large scale probabilistic properties of dynamical systems, including in particular their long time behaviour. The condition for a solution of the Liouville equation to reach an equilibrium state is that the dynamics must be mixing[42]; at the level of trajectories, it means that neighbouring trajectories diverge exponentially. An analogy may help. Imagine some perfume in a bottle in one corner of a room. The cap is removed and the vapours suffuse the room. Equilibrium is reached when the density of the perfume is uniform throughout the room – that is the equivalent to phase space equilibrium, when the probability distribution function no longer changes with time.

Since we can never know the true initial conditions for a real system (which arise as a consequence of whatever it was doing before we started to observe it), we are obliged to formulate the approach to equilibrium in probabilistic terms. Indeed, even a single trajectory associated with a given initial condition becomes increasingly inaccurate as time passes, since the exquisite sensitivity of the dynamics means that round-off errors accruing during the time integration of the equations of motion inevitably put the system on orbits other than the one it began on.

The lack of reproducibility thus stems primarily from the intrinsically chaotic nature of classical molecular dynamics. Other sources of uncertainty may play a role, many being potentially tractable; these include: the theory or the model used, the extent of convergence of the numerical method, the reliability of the software (which may not have been verified), the way the software is used, and so on[8,39]. We therefore focus on ensemble averaging, which is mandatory in statistical mechanics, for the convergence, reproducibility, reliability and uncertainty quantification of properties obtained from MD simulations. If we adjudge the system to be in a state of equilibrium, we can in addition perform time averaging, a procedure generally bereft of meaning out of equilibrium.

### Ensemble Method

Extensive studies we and others have performed in recent years[8,44,67–73,46–49,63–66] confirm that the most effective and reliable computational route to reproducible binding free energies of ligands to proteins using MD simulation can be achieved using ensemble methods. The same conclusion has been drawn from MD simulations in other areas, including studies on materials applications such as graphene based systems[74], on DNA nanopores using coarse-grained MD simulations[75], on rate parameter estimation for binding kinetics[76] and so on[77,78].

An ensemble approach employs a set of independent MD simulations, referred to as "replicas" both in statistical mechanics and within the uncertainty quantification domain, to obtain the required averages and associated. The key feature of such simulations is the use of ensembles and—for systems at equilibrium—time averaging. It is useful to recognise the stochastic nature of these simulations; it can be convenient to approximate the statistical properties of such ensembles as Gaussian random processes[42]. The requirement for the number of replicas and the temporal duration of the simulations depends mainly on the property of interest and the conformational space being sampled. Each replica needs to be simulated for long enough to sample the most relevant conformational space in order to evaluate one or more properties of interest. For free energy calculations using an end-point approach, for example, the stable binding states are the most relevant. There is no theoretical means to establish the number of replicas required to produce low errors from ensemble simulation: the criterion for ensuring convergence of the ensemble average is to establish the number $N$ of replicas required such that using $N+1$ of them makes no significant difference to the expectation values calculated. Of course, this can also be looked at another way, as amounting to a trade-off between the amount of computation one performs (which increases linearly with $N$) and the size of the error one is willing to tolerate (which reduces roughly as $N^{1/2}$). The computational costs can be optimised when a costs-to-accuracy ratio is carefully considered in a situation-specific and task-specific manner. Our studies show that starting from good initial structures, accurate and reproducible results can be achieved from an ensemble simulation consisting of 25 replicas with 4 nanoseconds production runs each using end-point approaches[8]. For a drug screening project, where a large number of compounds need to be evaluated, a coarse-grained workflow with a smaller number of replicas can indeed be useful to distinguish binders from nonbinders[79]. In such cases, replicas differ only in terms of the random number seeds selected to assign initial atom velocities. In other situations, it may additionally be necessary to randomly vary the initial atomic configuration, that is the spatial coordinates, as well[80]. These matters are discussed further below.

In general terms, the principles discussed above are applicable to all-atom MD and coarse-grained MD. The reduction in the number of degrees of freedom achieved by grouping several atoms into single particles or pseudo-atoms, as is done in coarse-grained MD, reduces the level of fluctuations in such systems. While this form of coarsening of the model's representation can typically lead to a decrease in accuracy, the benefits which accrue are an ability to study larger systems and for longer time periods; the reduced degrees of freedom also reduce the phase space that needs to be sampled. We find that smaller ensembles are required for coarse-grained MD than all-atom MD. This typically leads to an ensemble with around two-thirds the number replicas as compared to all-atom MD ensembles, as we have shown in recent work with graphene oxide dispersions[81] and DNA nanopores in lipid bilayers[82].

*Performing Ensemble Simulation*

The requirement to simply run a large ensemble of replicas may sound trivial but it comes with significant overheads in terms of managing the execution of simulations and collation of output data. It will be clear that using ensemble methods greatly adds to the computational cost of a study and to the wall clock time unless one has access to modern high performance computers which are equipped with large numbers of nodes, cores and often accelerators. In such cases, in the time it takes to run one simulation, one can produce the output for all of them. There is then a lot more data to handle, and process. The key step from the overall technical perspective is to bring all these output trajectory data together and then perform the analysis on the aggregate of all of this. This can be done in a number of ways, but the one we generally use is called a bootstrap error[83]. Given $N$ results from an ensemble of simulations, the bootstrap method involves calculating the distribution of means from resamples of size $N$ from that original results. Many resamples are taken, typically greater than 10,000, are made with replacement. If the original sample is representative of the true distribution, this method can provide error bounds or confidence intervals on any calculated value. The bootstrap error behaves similarly to a standard error; indeed, it is meaningful for quantities that have non-Gaussian distributions. This is of practical value in cases where we do not know the distribution of the quantity of interest.

Evidently, automation of some sort is necessary to manage the extra effort involved, and efficient sampling techniques are required to make these kinds of workflow possible. We have previously developed software to assist us in this task for the computation of binding free energies, the so-called "binding affinity

*Phil. Trans. R. Soc. A.*

calculator"[84]. This in turn led us to develop software for more general forms of uncertainty quantification, and to extend this to address verification and validation too. In particular, we have developed the VECMA Toolkit[30,85], as an open source, open development project which is allowing us to apply these methods much more widely, to address uncertainty quantification in a set of diverse domains. For example, the toolkit includes a python library called EasyVVUQ, which permits users to instrument their own codes with capabilities to perform UQ using a wide range of methods, including quasi-Monte Carlo (the method described here), stochastic collocation, and polynomial chaos[29,86].

### Statistical distributions revealed by ensemble simulation

The use of ensembles in molecular dynamics simulation only started to be systematically and routinely viable since the advent of the petascale era (that is, in a little over the past ten years), as a result of the vast increase in the number of nodes, cores and accelerators available on supercomputers. An instructive thing to do is to plot the frequency distribution of observables as it emerges from all the members of an ensemble, as this gives us an indication of the nature of the distributions we can expect. Figure 1 shows a set of examples of the data which typically emerges from these studies, and is drawn from work we have performed over the past few months[79].

It is clear from these plots that, while they are approximately Gaussian, they exhibit deviations from the standard bell curve expected on the basis that the variables are independent of one another, as one assumes in conventional statistics. Instead, we find that the distributions have a skewness associated with them, the asymmetry favouring the occurrence of values of the observable higher than the mean. The majority of the distributions have positive excess kurtosis, meaning they are heavy-tailed relative to a normal distribution. Our predictions of non-normal distributions prompted an investigation of the distributions of experimental binding free energies from a large number of independent measurements over time: these also exhibit non-normal properties (Ian Wall and Alan Graves, private communication (2020)). Caution is therefore required when statistical comparisons are made between predicted and measured thermodynamic properties, as it is widely assumed that these data are normally distributed. Such behaviour is at first sight unexpected, until it is recognised that these systems all display chaotic behaviour as well as long-range interactions[31]. A Gaussian distribution of free energy results can only be assumed for harmonic systems or transformations that can be approximated by linear response theory (see, e.g., Hummer *et. al*[87], Shirts and Pande[88], or König *et. al*[89]). Most free energy calculations are strongly influenced by anharmonic terms (e.g., van der Waals interactions), are not performed in a homogeneous environment (e.g., in a protein), or exhibit more than one dominant conformational substate[44]. The underlying nonlinearities in the dynamics are what accounts for both the presence of chaos and non-Gaussian statistics. The phenomenon is well known in turbulence: there it is caused by very long range hydrodynamic interactions mediated by energy dissipation. It is not anticipated from our experience of studying linear systems with short-range interactions at equilibrium.

That the equilibrium distributions arising in molecular dynamics should be non-Gaussian may appear surprising, given the accounts in most textbooks and lectures, which transfer themselves into research articles very readily. The reason for the presence of non-normal statistics in such systems at equilibrium comes from the fact that here too we are dealing with the infinite range interactions mediated by Coulomb forces. In a computer simulation of a closed system, such as the canonical or (N,V,T) ensemble, the molecular dynamics is driven by the existence of thermostats (and barostats in e.g. the (N,p,T) ensemble). The dissipation of energy within the system causes long-range correlations to be set up, which manifest themselves in the non-Gaussian nature of the statistics.

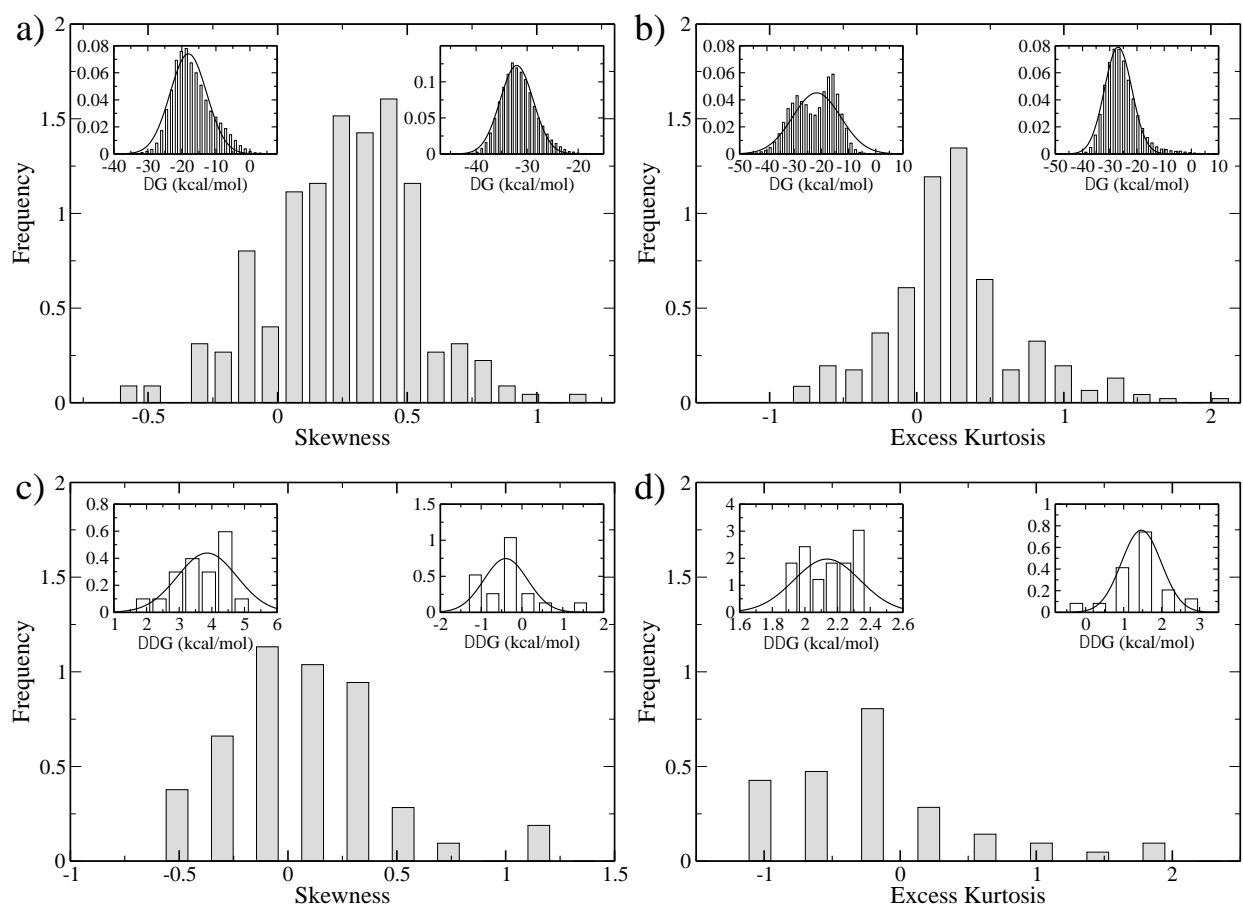

Figure 1. Molecular dynamics equilibrium distributions with long-range interactions are non-Gaussian. The main figures display the Fisher-Pearson coefficient of skewness (a, c) and the excess kurtosis (b, d) for distributions of predicted binding free energies (ΔG), or binding free energy differences (ΔΔG), using the ensemble-based molecular mechanics Poisson-Boltzmann surface area approach (ESMACS) (a, b) and thermodynamics integration (TIES) (c, d) approaches, respectively. The ESMACS results are obtained from 250 ligand-protein complexes, each with 25,000 frames accumulated from an ensemble simulation with 25 independent replicas. The TIES results include alchemical transformations of 50 pairs of ligands, from ensemble simulations comprising 20 or 40 replicas each. The inset shows distributions of binding free energies for two ligands, or ligand pairs, with the most negative and most positive kurtoses respectively. The best-fit Gaussian distributions are shown by black solid lines. See also Wan *et al.* [8]

*Phil. Trans. R. Soc. A.*

The practical implications of this discovery are important to apprehend. Non-normal distributions imply the occurrence of more "outliers" and an increase in the observation of so-called "rare events", making it harder to infer poor agreement between theory and experiment without far more data to pit down average behaviour. Naïve interpretation of correlation plots as implying poor agreement between experiment and theory when data do not magically cluster close to the 45-degree line need closer assessment; this is discussed further by Wan *et al.*[8]

## Bio-medical Simulation

There are various approaches to estimate the magnitude of the binding free energy (a measure of how strong the interaction is between a ligand and its target protein), based on different theories and approximations[90]. The "informatics" based approaches are, in the current era, usually the output of docking studies in combination with so-called "machine learning"[91]. The linear interaction energy (LIE) method[92] is an approximation of linear response theory; molecular mechanics Poisson-Boltzmann surface area (MMPBSA) and molecular mechanics generalised Born surface area (MMGBSA) methods[93] are based on invoking a continuum approximation for the aqueous solvent to approximate electrostatic interactions following all-atom molecular dynamics simulations; and, finally, so-called "alchemical" methods, including thermodynamic integration (TI) and free energy perturbation (FEP), are theoretically exact although in practice various approximations are made in their implementation. Binding free energy can also be calculated using approaches based on potential mean force; such calculations usually employ enhanced sampling approaches such as metadynamics, adaptive biasing forces or umbrella sampling[94]. The choice of which computational method to use is influenced by the desired accuracy, precision, time to solution, computational resources available, and so on.

### *Ensemble Method for End-point Approach*

End-point free energy methods allow one to explore configurational space in the protein–ligand bound and unbound states only, providing an efficient and accurate approach for the calculations of state functions such as free energies. MMPBSA and MMGBSA approaches are two commonly used end-point free-energy methods which require direct simulation of the two physical states. To generate the structures of these states, one can use a 3-trajectory (3-traj) approach in which separate MD simulations are performed for the ligand, apo (free) protein, and ligand-protein complex. Alternatively, one can use a 1-trajectory (1-traj) approach in which a single simulation is performed for the complex; a 2-trajectory (2-traj) variant allows for flexibility in the complex and one of the other two. The conformations for the ligand, protein and complex all being extracted from the complex simulation, the 1-traj approach makes use of an assumption that the conformations of the separated ligand and protein are similar to those of the complex. The assumption in the 1-traj approach is based on a lock-and-key hypothesis in which a substrate fits perfectly into the active site of an enzyme just like a key fitting into its lock. In the 1-traj approach, noise is significantly reduced as the energy terms are largely cancelled out. That assumption is questionable in many cases, however, as binding typically leads to conformational changes in both protein and ligands. The 3-traj approach does not use this assumption but the amount of noise is substantially increased when taking energy differences from three individual simulations. It is important to recognise that the 2- and 3-trajectory variants only become possible when ensemble-based methods are used, as extensive averaging is required to reduce the fluctuations present in individual trajectories, as is discussed further below.

### Errors in End-Point Approaches

The variation in the 1-traj free energy calculations based on ensemble simulations was investigated systematically by Sadiq et al.[46] and by Genhenden and Ryde[93] using MMPBSA and MMGBSA methods, respectively in 2010. The estimated free energies from two independent MMPBSA calculations of the same molecular system can vary by more than 10 kcal/mol in smaller molecule–protein complexes[47,49,93], and by up to 43 kcal/mol in larger and/or more flexible ligands bound to a protein such as the peptide–MHC (major histocompatibility complex) systems[48]. With the ensemble method, however, a meaningful ranking of binding free energies is generated. The 3-traj approach is able to address the role of the adaptation energy – the free energy associated with the conformational changes upon binding. The large adaptation energy, up to 39 kcal/mol, indicates that it is necessary to invoke the 3-traj approach for flexible small-molecule/protein binding. This has been confirmed by several subsequent studies[63,67,70] which show that incorporating the flexibility of the receptor and ligand improves the prediction of binding free energy ranking. There are also cases where incorporating flexibility does not improve the ranking[69], indicating that binding is mediated by a lock-and-key mechanism.

### Ensemble end-point simulations

To generate reliable, precise, and reproducible binding free energies from MMPBSA and MMGBSA approaches, we have proposed an ensemble based MD approach, named "enhanced sampling of molecular dynamics with approximation of continuum solvent (ESMACS)"[48]. This builds around the so-called MMPB(GB)SA method, including configurational entropy and free energy of association, but with important additional features to address reliability and reproducibility. Correctly accounting for entropic contributions is essential for reliably predicting binding free energies in cases where the ligands are diverse and/or flexible with many rotatable bonds. The contributions can be incorporated to the calculated free energies using normal mode (NMODE) approach or a variety of other options[70]. We have found a varying number of replicas may be required to achieve a desired level of precision; for many small molecule-protein systems, 25 replicas are typically required with 4 nanosecond production run for each replica, for ESMACS studies[48,49]. As already noted, the combination of the simulation length and the size of the ensemble provides a tradeoff between computational cost and precision. In collaboration with several pharmaceutical companies, we have used the ESMACS approach to investigate drug-like small molecules bound to therapeutic targets[67–70], and show that ESMACS is well suited for use in the initial hit-to-lead activities within drug discovery.

### Ensemble Method for Alchemical Approaches

The alchemical approach calculates relative binding free energies between two physical states which are linked by an "alchemical" path. A series of nonphysical steps are involved in the path. The two physical states can be a protein binding with two ligands, or a ligand binding with wild-type and mutant proteins. Along the alchemical path, some atoms change their chemical identities – appearing, disappearing or alchemically transforming from one to another. Although the alchemical free energy methods are formally exact and general, the possible large uncertainties and expensive computational cost limit the domain of application: they are applicable mainly to estimating small relative free energy changes for structures (drugs or proteins) which involve relatively minor (perturbative) variations. The alchemical method can also be used to calculate absolute binding free energies[65]. It is the equivalent of the relative free energy calculation when one of the ligands involved is replaced by nothing, and thus face even more demanding challenges to achieve convergence.

Free energy calculations using such alchemical methods had rarely been used seriously in drug development projects until recently when Schrödinger Inc. released their "FEP+" simulation software for relative free energy calculations[95]. With the improved methodology, much of which is proprietary and thus not available for assessment, and the use of graphical processing units (GPUs), FEP+ has made a significant impact in the pharmaceutical industry within its domain of applicability[3], although further evaluation is still needed on its accuracy and precision[44,63,65]. From the perspective of this study, however, it is interesting to observe that the methodology advocated is decidedly based on use of "one-off" simulations, so that any attempt to provide uncertainty quantification is entirely lacking here.

### Errors in alchemical calculations

As in many other approaches, an alchemical calculation certainly generates random errors, and very likely systematic errors too. As we have stated above, we need to correctly handle the stochastic errors before we can reliably estimate the possible systematic errors. A survey of publications and binding databases shows that the binding affinities are in a range between -6.5 and -15.2 kcal/mol for most of interesting biomolecular ligands[96].

Molecular simulations aim to predict free energies accurately, with an error ~ 1 kcal/mol, for small molecules to their target proteins. Thermodynamic integration can produce significant differences from individual replica simulations, up to 1.58 kcal/mol and ~7 kcal/mol for relative and absolute binding free energies, respectively, for the cases tested using 5 independent simulations[65]. These simulations vary only in their initial velocities which are randomly drawn from a Maxwell–Boltzmann distribution. Similar results are obtained from multiple runs of FEP+ calculations, in which up to 3.9 kcal/mol variations have been observed from 30 independent simulations, much larger than the MBAR (multistate Bennett acceptance ratio) errors reported for individual FEP+ calculations[63].

When random errors are handled correctly, it is possible to identify the systematic errors intrinsic to these simulation methods, provided due attention is paid to the way in which errors, both theoretical and experimental, are handled. Systematic errors tend to shift all of the measurements/predictions for the same target systems with the same setup in the same direction from their real values. Different techniques have been applied in alchemical free energy simulations to enhance sampling in the hope of reducing errors in the calculations. Widely used techniques include accelerated sampling methods, such as replica exchange with solute tempering (REST2)[97], and free energy estimators, such as MBAR[98]. However, our studies paying careful attention to uncertainty quantification show that these techniques offer no guarantee of improving the accuracy of the predictions[44]. Indeed REST2, as used in FEP+, appears to generate a significant systematic underestimation of free energy differences, which degrades monotonically with duration of simulation[63].

### Ensemble alchemical simulations

Alongside ESMACS, we have developed an ensemble based approach called TIES (thermodynamic integration with enhanced sampling)[66]. TIES employs an ensemble of independent MD simulations in combination to yield accurate and precise free energy predictions. It quantifies and reduces the random errors, making the results precise and reproducible. This approach also makes it possible to distinguish the systematic errors, and to interpret the results correctly.

As one example among many, the application of TIES to protein mutations provides insights underpinning the impact of the gatekeeper mutation of the FGFR-1 kinase on drug efficacy[65]. Using an ensemble based approach, we were able to quantify the uncertainties in the free energy calculations, and to compare the performance of different software and hardware for the calculation of the same free energy changes[63]. Ensemble approaches like TIES provide a reliable, rapid and inexpensive method for uncertainty quantification applied to both relative and absolute binding free energy calculations using alchemical methods[63,65].

### Materials Simulation

*Phil. Trans. R. Soc. A.*

Predicting the properties of modern advanced materials typically requires understanding the structure and the dynamical processes on an atomistic level[77,99]. Many large-scale, macroscopic, engineering properties can be modelled using methods taken from continuum mechanics, such as the finite element method. However, diffusive processes, self-assembly, structural degradation, surface/interface characteristics, and many other quantities of interest to modern materials scientists and engineers, are all heavily influenced if not controlled by dynamical processes on the scale of atoms and molecules. This is particularly clear in the case of nanocomposites, in which one has to deal with a polymer matrix in which is embedded a nanomaterial such as graphene (and its oxide), carbon nanotubes, clays and such like[74,81]. Graphene and other so-called two-dimensional materials have one dimension which is of the order of nanometers, and thus just one or a few atoms thick, yet they impart dramatically enhanced large-scale materials properties in the composites they produce. In such circumstances, it is clear that one must use MD as part of the range of techniques available for studying such complex systems.

MD techniques are uniquely equipped to explore processes that occur on time and length scales of nanometers to microns, and nanoseconds to microseconds. For complex systems, especially those with anisotropic structures on the nanoscale, such as "soft matter systems", MD has proven useful for predicting the nanoscale structure and material properties. In the development of new materials, the chemical constituents are often among the first known aspects of the system, and the subsequent time necessary for developing useful applications is spent optimising the fabrication and processing for engineering tests. MD can often help to reduce if not remove many of these practical barriers and assess a material's suitability for a given purpose based only on its atomistic structure. Indeed, there is substantial interest in many areas of materials design in virtual testing using computer simulation to speed up the process from concept to real-world implementation, which currently takes of order twenty years, at a cost of many billions of dollars[100]. The challenge then becomes providing reliable computer based "*in silico*" predictions which reduce the need for expensive and time-consuming experimental work. This puts a premium on providing tight error bars since these furnish a key measure of the confidence with which we can accept modelling results and use them to guide experimental work.

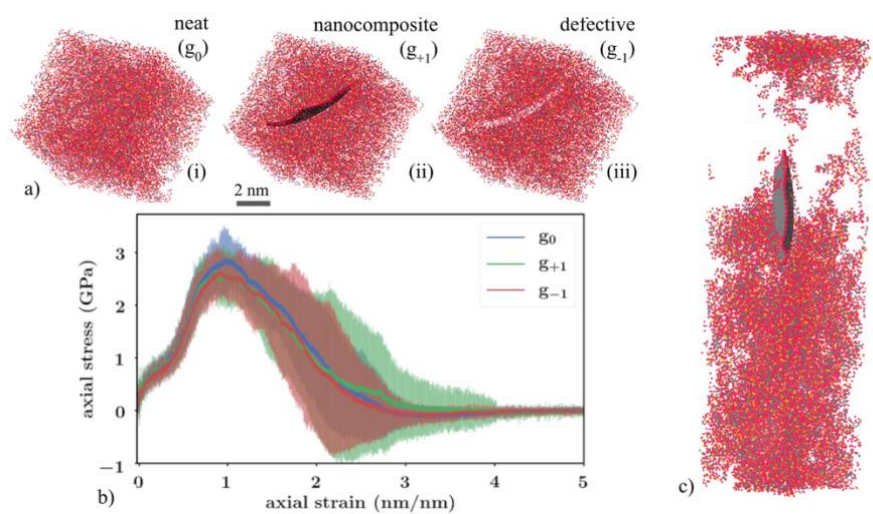

Figure 2: Measuring the toughness of different materials with a reactive forcefield. (a) Three structures: (i) neat epoxy polymer; (ii) epoxy-graphene nanocomposite; and (iii) epoxy polymer with a defect are strained uniaxially. (b) The stress-strain curves are shown in the plot; lines indicate the average of six replica simulations while the shaded regions correspond to the standard deviations at each strain. While each replica varies, the ensemble average shows the three materials behave similarly. (c) Displays a snapshot from sample (ii) at the point of fracture.

## Materials property prediction

By way of example, consider trying to predict a material's stiffness using molecular dynamics (see Figure 2)[101]. This is one of the most common applications of MD in materials science. A uniaxial stretch of such system is a fairly trivial simulation to perform; however, it poses several fundamental questions about the certainty we can have in a result of an MD simulation. As discussed above, several systematic uncertainties exist with this technique which are hard to quantify[102], the most glaring example of which is the choice of force field used to represent the material[103]. The often quoted limitations of MD—such as finite size/time effects or structure generation—are also systematic errors. Using appropriate workflows we can quantify these effects to produce accurate results[104].

MD simulations in the condensed phase are typically performed by imposing periodic boundary conditions in all three spatial directions, which means that we only expect to simulate a comparatively small simulation cell to approximate the bulk properties. The size of this simulation cell has many implications for computational cost but, more importantly, the reliability of the scientific results it furnishes. Finite size effects and fluctuations can be expected to affect the outcome of a simulation.

To measure the Young's modulus (YM) of a material system, the pressure exerted along one axis is sampled before and after (or during) imposition of a small strain. Since the instantaneous pressure of a molecular dynamics simulation can fluctuate by several GPa, it is necessary to average this value over a long sample period to measure the change in pressure due to an applied strain. In a recent study[29], we considered an epoxy resin system, a thermosetting polymer. The investigation quantified the effect of specific MD parameters on the measured Young's modulus, including the system size, starting velocities, and polymer generation random seed.

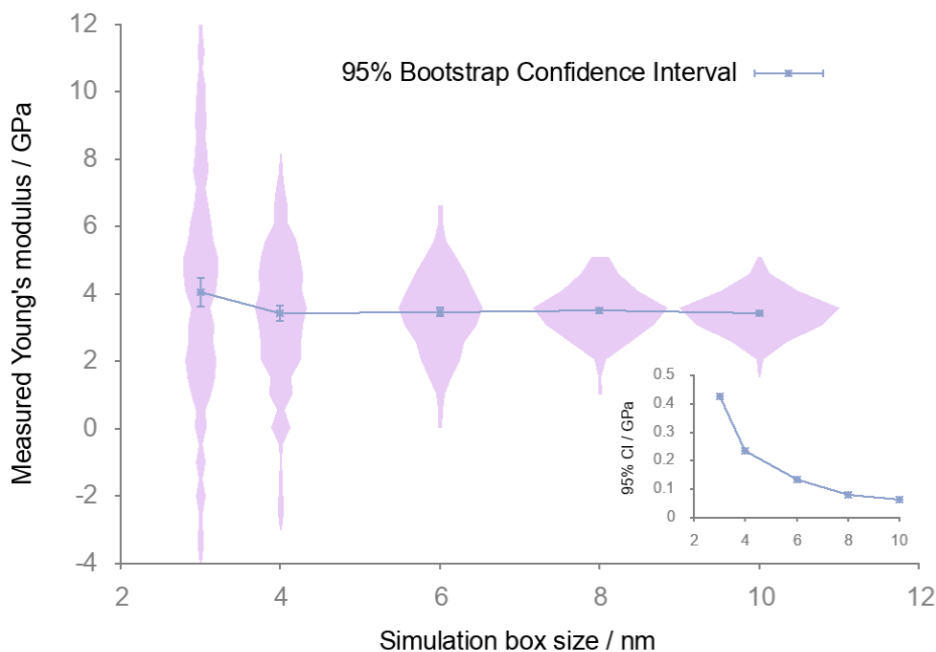

Figure 3: Young's modulus of an epoxy resin measured with different simulation sizes. Each point is the average of 300 simulations, which make up the pink histograms for each box size. The 95% bootstrap confidence interval for increasing ensemble size is shown in the inset plot at the bottom right.

We found that the mean YM of an ensemble of simulations is independent of simulation size but below a box of size 4 nm there is a finite size effect which makes the system artificially stiffer (Figure 3). This effect only became evident after performing 300 replica simulations at each box size, with mean YM and standard deviation $3.4 \pm 1.9$ GPa. The smallest box size gave a distribution of YMs with a significant skewness of -0.8; as the box size increased the skewness tended to zero. The distribution is effectively Gaussian because there are no long-range interactions present. The analysis was greatly facilitated by use of the EasyVVUQ software[29]. From this, it should be clear that one single measurement, at the low strains imposed here, is wholly inadequate to measure this property reliably.

Another benefit of running ensemble simulations is that one can perform sensitivity analysis and thereby learn about the variance in properties arising due to different input variables and parameters. By applying the law of total variance, our study showed that the expected variance due to the polymer network generation was equal to that due to the starting velocities of the atoms in the simulation. In other words, the exact connectivity of monomer units was inconsequential, provided that the same crosslink density is achieved. However, one single simulation of such a polymer, is insufficient; the aforementioned ensemble of 300 replicas was required in order to be 95% confident of the size effect seen above.

These approaches are broadly applicable. Running ensembles that give statistically significant results not only allows us to efficiently sample more phase space, therefore increasing the accuracy in the result, but provides much more information that single simulations would not be able to tell us.

Ensemble techniques are necessary for exploring phase space whenever molecular dynamics simulations are used. Alfè *et al.* explored the kinetics of phase transitions in superheated metals caused by nucleation processes[105,106]. The time taken for a superheated metal to melt in this manner is generally unknown: by simulating an ensemble of 350 replicas, differing only in the starting velocities of the atoms concerned, they were able to make accurate predictions of the system's kinetics. They found that nucleation rates are highly dependent on the details of nanoscale behaviour. Here too, finite size effects were isolated and corrected for in reporting reliable results.

### Generating structures

It is often forgotten that the starting structure can itself be a major source of uncertainty in MD simulations; it is as true for materials as it is in biomedical simulation. The most straightforward way of achieving variation in an ensemble is to use different random numbers to seed the initial atomic velocities; however, generating different starting configurations should also be considered as we did when performing an MD study of the Nobel prize-winning discovery of graphene by peeling off atomic layers from graphite[80]. The process of building the initial structure and coordinates of a system is not trivial and can often be the most time-consuming step in such research work: it may involve non-trivial polymer chain building for synthetic and biological macromolecular structures, the details about the microstructure of nanocomposites, and so on. The initial state of an MD simulation is inevitably artificial and therefore not itself a representation of the system of interest. Initial states must be built such that, for example, if one wishes to study the equilibrium state, it will not take an inordinately long time to reach that state by MD simulation. The starting structure must be sampled sufficiently and carefully assessed to check that unphysical starting conditions do not influence the production stage of a simulation.

Polymer systems are manifestly difficult to build while diffusion in high molecular weight polymers can be extremely slow, so entanglements and anisotropic structures must be built carefully as good starting points. Numerous techniques exist to do this. Diffusive processes are so slow that 'sampling the phase space' with one long simulation would be extremely expensive using normal molecular simulation approaches and inaccurate into the bargain; instead several structure generations are essential to be confident in a result.

Graphene oxide poses a different problem as its structure is that of an amorphous crystal[107]. Oxygen containing groups are present across the surface of graphene with some random distribution; in the presence

*Phil. Trans. R. Soc. A.*

of other materials, the precise distribution of these groups may influence their interaction. In the case of a graphene oxide dissolved in a polymer melt, it is not sufficient to build one structure and generate an ensemble with different initial velocities; instead one must generate several graphene oxide structures to understand the system.

*Forcefield critical properties*

The errors caused by inaccurate force fields can be completely catastrophic for the physics under investigation. A stable inaccuracy in a force field may give a binding energy within some error of the true value, but a more significant inaccuracy may result in divergent dynamical and/or structural regimes[108]. For example, when predicting the structure of crystals, a flawed force field may never produce certain crystal arrangements. Lennard-Jones forcefields are known to underestimate the friction between layered materials[109]. Sinclair *et al.*[74] found that the spherical symmetry of these potentials is too gross an approximation whilst simulating graphene bilayers so a new forcefield was developed. In experiment, graphene flakes are observed to show superlubric behaviour when propelled across a graphitic surface. Propelling flakes in this way is a chaotic process (in the technical sense that it is highly dependent on the initial conditions). In order to achieve an acceptable error which was comparable with experimental data on frictional properties, ensembles of 40 replicas were required. The distance travelled by a flake seems to follow a lognormal distribution, with a Fisher-Pearson coefficient of skewness of 1.4.

## Generating Actionable Predictions

In order for the predictions from computational science to inform costly and consequential decisions for real-world problems, it is vital that they are accurate, reproducible, and accompanied by uncertainty quantification. Speed too is of critical concern, in order for predictions to be actionable — that enables decision makers to take appropriate actions in a certain period of time. Aerospace manufacturers are keen on the concept of virtual certification in order to reduce the time to market, along the way from concept to implementation. Approaches and tools have been proposed to perform uncertainty analyses in real-world practice[14]. Emulator based methods[14] and the test harness[110] are used to analyse sensitivity and uncertainty, to evaluate scientific software and to calibrate complex computer simulators. Practical recommendations have been made on the validation and reproducibility in the field of scientific research in general[111] including molecular dynamics[5,39]. Uncorrelated data need to be collected in sufficient quantities; autocorrelation analyses can be used to better understand if a time series represents an equilibrated system[111], but caution must be exercised as longer autocorrelation times may be uncovered if other relevant free energy minima are revealed[45].

In computer-aided drug discovery, different levels of uncertainties are needed in various stages, including searching for potential small molecule hits, identifying promising leads, and optimising leads for further evaluation. Concentrating on the objects of uncertainty can give us an appreciation of how decisions can be made based on uncertainty quantification and cost-benefit analyses. The Accelerating Therapeutics for Opportunities in Medicine (ATOM) project (https://atomscience.org/), funded by the Department of Energy in USA, is predicated on the aim of being able to go from concept to clinical trials for new drugs within 12 months. This is a very large project involving many partners, from academia, industry and US national laboratories. Being able to rely on the results of *in silico* predictions will undoubtedly turn these industries on their head, by eliminating large amounts of laboratory based testing. Knowledge of the uncertainties attaching to decisions is critical here.

In the field of medicine, where patient safety is at stake, it is extremely important to have an uncertainty quantification discipline for risk management[112]. We are active in the context of clinical decision making for personalised medicine and have discussed these requirements for many years (see e.g., Groen *et. al.*[113], and Manos *et. al.*[114]). Reproducibility and uncertainty (indeed, more completely, VVUQ) are of central concern for the predictions to be deemed actionable: for uptake of our *in silico* approach in medical and clinical contexts, regulatory authorities will demand procedures which are fully certified in this sense. This can be done through careful control of the uncertainties in the calculations employed, which rest heavily on molecular dynamics computation.

Generating actionable predictions requires a high level of automation which can be achieved through a powerful combination of software and hardware, making calculations immediately scalable for industrial and clinical applications. To make predictions which are fast enough for actionable decision making, urgent priority must be given to such calculations on high performance computers. Our requirements for on demand access to large scale computing resources for such purposes are well known and have informed numerous initiatives across the world to provide appropriate access mechanisms for medical and clinical research. This is part of the core business of the Computational Biomedicine Centre of Excellence (www.compbiomed.eu) which has innovation and sustainability at its heart. As a matter of fact, the EasyVVUQ software which we have introduced as an open source toolkit, and as part of the VECMA Toolkit[30], is based on its original applications to MD as discussed here. It is now being applied widely in many fields of concern to this theme issue of *Phil Trans A*, from fluid dynamics to climate prediction and fusion energy research. As such, it can be reasonably considered that the study of uncertainty quantification in MD is now beginning to have an impact within the wider general field of VVUQ.

## Conclusions

*Phil. Trans. R. Soc. A.*

Ensemble molecular dynamics simulation provides us with a powerful methodology that enables us to connect ergodic theory and uncertainty quantification, and to obtain reproducible results from simulations in a systematic and theoretically well-grounded manner. As evidenced in the case of ligand-protein binding free energy predictions, ensemble simulation-based approaches yield statistically robust, precise and reproducible, hence reliable results. Using ensemble methods, the errors in predictions can be systematically controlled, amenable to further reduction by increasing the number of replicas in an ensemble and in propitious circumstances too by extending the length of such simulations. Ensemble approaches are scalable: for example, they permit hundreds to thousands of binding affinities to be calculated per day, depending on the computing resources available. Computing capabilities are set to increase as the exascale era heaves in to view. In the near future, rapid, accurate and reliable predictions of materials properties may emerge that can be exploited in the aerospace and automotive industries; free energy prediction at high throughput will assist physicians in clinical decision making and medicinal chemists in directing compound synthesis in a routine manner. In sectors such as these, virtual certification and regulatory approval for the use of *in silico* methods will depend critically on the application of rigorous uncertainty quantification along the lines we have described.

**Authors' Contributions**
All authors contributed to the concept and writing of the article.

**Competing Interests**
The authors have no competing interests.

**Funding Statement**
We are grateful for funding from the UK EPSRC for the UK High-End Computing Consortium (EP/R029598/1), from MRC for a Medical Bioinformatics grant (MR/L016311/1), the European Commission for the CompBioMed, CompBioMed2 and VECMA grants (numbers 675451, 823712 and 800925 respectively) and special funding from the UCL Provost.

The authors made use of the BlueWaters supercomputer at the National Center for Supercomputing Applications of the University of Illinois at Urbana-Champaign (https://bluewaters.ncsa.illinois.edu), access to which was made available through the aforementioned NSF award; and the Titan and Summit supercomputers at the Oak Ridge National Laboratory, supported by the Office of Science of the U.S. Department of Energy under Contract No. DE-AC05-00OR22725. The authors acknowledge the Gauss Centre for Supercomputing and the Leibniz Supercomputing Centre in Garching, Germany, for providing access to SuperMUC-NG (https://www.lrz.de/services/compute/) through a Large Scale Gauss award (project number pn98ve).

**Acknowledgments**

PVC is grateful for many stimulating conversations with Dario Alfè, Bruce Boghosian, Wouter Edeling, Alfons Hoekstra, Peter Sloot and Sauro Succi. We thank Agastya Bhati for his help in finalising the manuscript.